\documentclass[12pt]{iopart}

\usepackage{changepage}

\usepackage[utf8x]{inputenc}

\usepackage{textcomp,marvosym}

\usepackage{cite}

\usepackage{nameref,hyperref}

\usepackage[right]{lineno}

\usepackage{microtype}

\usepackage[table]{xcolor}

\usepackage{array}

\usepackage{graphicx}
\usepackage{float}
\usepackage[english]{babel}
\newcommand{\newpara}
    {
    \vskip 0.2in
    }

\begin{document}
\vspace*{0.2in}

\begin{flushleft}
{\Large
\textbf\newline{Coordination of size-control, reproduction and generational memory in freshwater planarians} 
}
\newline

\author{Xingbo Yang \textsuperscript{1*}, Kelson J. Kaj \textsuperscript{2*}, David J. Schwab \textsuperscript{1}, Eva-Maria S. Collins \textsuperscript{2,3}}
\address{1 Department of Physics and Astronomy, Northwestern University, Evanston, Illinois, USA\\2 Department of Physics, University of California San Diego, La Jolla, California, USA\\3 Division of Biological Sciences, University of California San Diego, La Jolla, California, USA}
\vspace*{0.2in}
* These authors contributed equally to this work.

Correspondence: david.schwab@northwestern.edu or emscollins@physics.ucsd.edu




\end{flushleft}
\section*{Abstract}
Uncovering the mechanisms that control size, growth, and division rates of organisms
reproducing through binary division means understanding basic principles of their life
cycle. Recent work has focused on how division rates are regulated in bacteria and yeast,
but this question has not yet been addressed in more complex, multicellular organisms.
We have, over the course of several years, assembled a unique large-scale data set on the growth and asexual reproduction
of two freshwater planarian species, \textit{Dugesia japonica} and \textit{Dugesia tigrina}, which
reproduce by transverse fission and succeeding regeneration of head and tail pieces into
new planarians. We show that generation-dependent memory effects in planarian reproduction need to be taken into account to accurately capture the experimental data. To achieve this, we developed a new additive model that mixes multiple size control strategies based on planarian size, growth, and time between divisions. Our model quantifies the proportions of each strategy in the mixed dynamics, revealing the ability of the two planarian species to utilize different strategies in a coordinated manner for size control. Additionally, we found that head and tail offspring of both species employ different mechanisms to monitor and trigger their reproduction cycles. Thus, we find a diversity of strategies not only
between species but between heads and tails within species. Our additive model provides two advantages over existing 2D models  that fit a multivariable splitting rate function to the data for size control:
Firstly, it can be fit to relatively small data sets and can thus be applied to systems where available data is limited. Secondly, it enables new biological insights because it explicitly shows the contributions of different size control strategies for each offspring type.

\section*{Introduction}
{\color{black}Freshwater planarians are famous for their regenerative prowess. Somatic pluripotent stem cells enable them to regenerate from even a small piece of the original animal~\cite{Wagner2014,Baguna2012,Rink2013}. Their ability to regenerate is closely linked to their ability to reproduce asexually by binary fission, whereby the worm tears itself into two pieces (head/tail), which subsequently regenerate the missing body parts to become full planarians. How they do this and which parameters control their reproduction cycles, however, remain unknown. This is largely because fission is rare, with division rates on the order of 1/month or less \cite{Dunkel2010,Quinodoz2011,Thomas2012,Carter2015}, making it challenging to accumulate sufficient data for statistical analysis within a reasonable time frame.
\newpara
Using our SAPling (Scan-Add-Print) barcoding database system \cite{Thomas2011} we have, over the course of several years, accumulated a unique large-scale data set on planarian asexual reproduction. We tracked the lineages and waiting times of thousands of individual planarians from three species originating from three different continents \cite{Carter2015}, \textit{Schmidtea mediterranea} (S-worms), \textit{Dugesia japonica} (J-worms), and \textit{Dugesia tigrina} (G-worms)\cite{Dunkel2010,Quinodoz2011,Thomas2012,Carter2015}. For a subset of the data (hundreds of individuals), we furthermore imaged the planarians at division to record birth size, division size, and the proportions of head and tail pieces. Together, these data fully describe an individual's growth and reproductive behavior on the organismal level.
\newpara
Because all species use the same mode of reproduction, i.e. tearing themselves into two offspring, we expected to find similarities between their reproductive behaviors. Indeed, we found that \cite{Carter2015}: (1) Division is asymmetric for all species, with head pieces generally being larger than tail pieces. (2) The division rates of head and tail offspring, i.e. planarians originating from the head or the tail piece in a division, differ substantially. Head offspring generally divided more frequently than tail offspring. This is a direct consequence of size differences and the fact that the tail piece needs to regenerate more complex structures (such as the brain and eating machinery), and thus can initially not eat and grow as quickly as the head offspring \cite{Thomas2012, Carter2015}. (3) All species exhibit a ``generational memory'', i.e. their reproductive behavior is not only influenced by their type (head/tail) but also by their history. In particular, it matters for an individual's reproductive behavior what type parent (head/tail) it had and whether the parent fissioned once and then fully regenerated, or fissioned multiple times in a row before fully regenerating (``fragmentation'') \cite{Dunkel2010,Quinodoz2011,Thomas2012,Carter2015}. Given these commonalities, we were surprised to find that the reproductive strategies employed by the three species are distinctively different. S-worms primarily fragment, thus producing many small offspring in a short period of time and then wait a long time to repeat the process. J-worms divide comparatively more frequently and primarily fission, producing fewer but larger, roughly equally sized offspring. G-worms fission at the highest frequency of the three species and produce small tails while keeping large heads \cite{Carter2015}. We interpreted these differences in strategy and resource allocation as arising from adaptation to different environmental challenges in their geographically distinct natural habitats \cite{Carter2015}.
\newpara
In the light of these data, we now ask the question of how an individual planarian controls its size and reproductive behavior. {\color{black}This fundamental question has been asked and addressed for bacteria and yeast \cite{John1983,Talia2007,Skotheim2012,Amir2014,Araghi2015,Jun2015,Jun2016,Campos2014,Iyer2014, Banerjee2016}. Three phenomenological size control strategies have been identified and termed as sizer, adder, and timer, corresponding to division cued by an organism's size at division, its accumulated size between divisions, and the time passed since its last division (``waiting time''), respectively. The strategies are quantified by a splitting rate function of the corresponding variable \cite{John1983,Skotheim2012,Araghi2015,Jun2015,Jun2016,Campos2014,Amir2014,Talia2007}. The splitting rate function can be obtained by integrating the experimental probability distributions of the division size, added size and waiting time to yield an empirical estimation of the reproductive strategy. Mixed strategies can also be studied by constructing splitting rate functions of more than one variable, yielding for example, sizer-adder and sizer-timer models \cite{Osella2014}. These models are accordingly referred to as ``2D models''.

One caveat is that the fitting procedure generally requires large amounts of experimental data. How much data is necessary depends on the complexity of the model, therefore making the identification of the reproduction strategy data-limited. Recent works using microfluidics experiments to study \textit{E. coli} reproduction have generated sufficiently large data sets to overcome this difficulty and shown that \textit{E. coli} reproduction dynamics can be accurately described by the adder principle~\cite{Araghi2015,Jun2015,Jun2016,Campos2014} or a mixed strategy of sizer-timer mechanism \cite{Osella2014}.}
\newpara
{\color{black}The existing single cell models provide a natural starting point for the study of asexual planarian reproduction and size control. While the biology and the time-scales of reproduction are quite different due to the multi-cellular nature of planarians, both planarians and single cell organisms produce two offspring in each division cycle. However, as we show here, data limitations due to the long waiting times and the experimentally observed generational memory of planarians prevent the fitting of a full 2D model.

We therefore introduce a new additive model that mixes multiple size control strategies based on planarian size, growth, and waiting time data. The strength of this additive model is three-fold: Firstly, it can be fit to the experimental data, which is less abundant compared to published single cell data. Secondly, it successfully captures the generational memory and makes accurate predictions on the generational dynamics. Thirdly, it straightforwardly quantifies the proportion of each strategy in the mixed dynamics. We find that J-worm tail offspring are best described as pure sizers, G-worm heads as pure adders, and J-worm heads and G-worm tails are equally well described by mixed “sizer-adder” or “sizer-timer” models.}

This diversity of strategies, not only between species but between heads and tails within species, points toward intrinsic physiological differences of heads and tails. A major difference is clearly that at birth, head offspring have a brain and tails don't. However, the finding that neither heads nor tails of the two species studied here share the same strategy, suggests that the explanation is not that simple. Relating the observed behaviors to their biological origins remains a future challenge. The quantitative analysis on the mechanisms of size control on the organismal level presented here will help guide mechanistic studies aimed at dissecting the molecular components regulating growth and reproduction in asexual planarians.

\section*{Materials and Methods}
\subsection*{Planarian Maintenance}
Two species of freshwater planarians were used in all experiments: \textit{Dugesia japonica} (J) and \textit{Dugesia tigrina} (G). J-worms were asexual, clonal strains from established lab colonies in China. G-worms were purchased from Ward's Science and had been caught in the wild in Wisconsin, U.S.A. No eggs were found among G-worms during the course of the experiments, so we assume that they only reproduced asexually. Worms were kept in dark at 20$^{\circ}$ in Panasonic incubators, except to be fed once per week, cleaned twice per week, and imaged twice per week. Planarians were fed organic beef or chicken liver purchased from a local butcher.

\subsection*{Isolated Worm Experiments}
Two J-planarians and two G-planarians were randomly selected to be ``founder worms" for establishing family lineages, as described previously \cite{Carter2015}.  These worms were placed in individual petri dishes (100mm diameter x 20mm height, Fisher Scientific) containing 25 mL of planarian water.  Asexual reproduction events were recorded with a time resolution of 2-3 days.  When a reproduction event occurred, individuals were separated to individual petri dishes with barcodes holding information about the individual's family birth date and family history.  This information was recorded in the SAPling database developed in \cite{Thomas2011}.  A total of 8,568 J- worms and 2,492 G-worm reproductive events were observed in a period of 7 years and 3 months.

\subsection*{Growth Rates}
20 J-worms and 23 G-worms were randomly selected to be ``founder worms" for measuring individual growth rates.  Growth rates were measured for multiple worms in each family, both heads and tails.  These planarians were kept under the same conditions as worms in the Isolated Worm Experiments and imaged twice a week. A total of 149 full growth rates were collected over a time period of 9 months.

\subsection*{Image Acquisition and Area Analysis}
Image acquisition and area analysis was performed as described in \cite{Carter2015}. In brief, individual planarians were imaged with a Basler A601f CCD camera (Basler AG, Ahrensburg, Germany) using a Leica stereo microscope (Leica Microsystems, Wetzlar, Germany) and Basler Pylon Viewer (Basler AG, Ahrensburg, Germany).  Since planarians are essentially flat, size was estimated by area instead of volume. Areas were obtained using ImageJ (U.S. National Institute of Health, Bethesda, MA) and a custom MATLAB (MathWorks, Natwick, MA) script.  Area of worm at division was obtained by summing the areas of all offspring with a time resolution of 2-3 days.

{\color{black}\subsection*{Simulation}
{\color{black}We simulate the asexual reproduction dynamics of planarians based on the 2D model and the additive model presented in the main text, both of which incorporate mixed sizer-adder or sizer-timer strategies.} We begin with a single founder worm as in the experiment and set a time step of $dt=1$ corresponding to 1 day in real time, and for each time step we keep track of the worm size, added size and waiting time (RWT). We calculate the worm size $s$ according to an exponential growth law with growth rate $\lambda$ drawn from the experimental distribution using
\begin{equation}
\label{eqn:growth_law}
s = A\exp(\lambda \tau),
\end{equation}
where $s$ is the worm size, $A$ is the birth area, $\lambda$ is the growth rate and $\tau$ is the RWT between two divisions. The added area $\Delta$ can be computed as $\Delta = s-A$. Then we use the division probabilities as defined in Eqs (\ref{eqn:2D_division_sizeradder})-(\ref{eqn:2D_division_sizertimer}) for the 2D model and Eqs (\ref{eqn:division_sizer_adder})-(\ref{eqn:division_sizer_timer}) for the additive model to determine whether a division happens or not by drawing a random number from $0$ to $1$. The division probabilities are obtained by integrating the probability distributions of Teilung size, added size and RWT. When a division occurs, the parent worm divides into a head (H-) worm born from the head piece and a tail (T-) worm from the tail piece. The division is generally asymmetric, and the proportions of head and tail pieces are drawn from the corresponding experimental distributions. We terminate the simulation when we reach the same number of worms as in the experiment and record birth size ($A$), Teilung size($s$), added size($\Delta$), RWT ($\tau$), and lineage of all the worms for data analysis.

\subsection*{Data Analysis}
Out of all the reproductive events observed, we recorded birth area, division area, RWT, and lineage for 378 J H-worms, 322 J T-worms, 151 G H-worms and 59 G T-worms across two clonal families for each species. All worm data was organized and analyzed using custom MATLAB (MathWorks, Natwick, MA) scripts, and the results are shown in Fig.\ref{Fig1}-\ref{Fig3}. We run $500$ simulations for the 2D model and $500$ simulations for each pair of $\alpha$ and $\beta$ of the additive model for both sizer-adder and sizer-timer strategies. We fits the simulation data to obtain various slopes with reference to Fig.\ref{Fig1} and calculate the mean Teilung size, birth size, added size, and RWT for worms of different generations with reference to Fig.\ref{Fig2}. The former is called the population data and the latter the generation data. For the additive model, we plot histograms out of the $500$ simulations for each pair of $\alpha$ and $\beta$ and fit a Gaussian distribution to the histogram to obtain the probability distribution of the population and the generation data, which is the model prediction. We extract the log likelihood as defined in Eq (\ref{eqn:likelihood}) by evaluating the likelihood of observing the experimental population and generation data given the model parameters. The optimal model parameters $\alpha$ and $\beta$ are determined through maximum likelihood estimation (MLE). We emphasize that the observables that we are evaluating the likelihood on, i.e. the population and the generation data, are not the data we use to fit the splitting rate functions. Therefore, the log likelihood serves as a metric to determine the predictive power of the model. We have tested our inference scheme using artificial data generated by the additive model with known $\alpha$ and $\beta$. We correctly recovered $\alpha$ and $\beta$ by maximizing the log likelihood function for both sizer-adder and sizer-timer model as shown in Fig.\ref{Fig7} in the SI. However, we are unable to distinguish the optimal sizer-adder model from the optimal sizer-timer model. This performance equivalence of the two mixed models has previously been reported in \textit{E.coli} reproduction dynamics \cite{Osella2014}.}

\section*{Results and Discussion}
\subsection*{Planarian reproductive behaviors from experimental data}

{\color{black}We have previously identified two modes of asexual reproduction in planarians: fission and fragmentation \cite{Quinodoz2011}. In a fission event, the planarian splits into a head and a tail piece, which subsequently regenerate into two new worms. In a fragmentation event, the worm divides multiple times before complete regeneration occurs, thus producing a head, tail and one or more trunk pieces \cite{Quinodoz2011}. Fragmentations are the main mode of reproduction for S-worms, but are rarely observed for J- and G-worms \cite{Carter2015}. Because trunk offspring are not captured in the binary division framework, we focus on J- and G-worms in this study and exclude any fragmentation from our analysis.

Each reproductive event (``Teilung''  \cite{Carter2015}) is thus a fission, and we will use the terms division, fission, and Teilung interchangeably in this paper. We refer to the worm that develops from the head offspring as a (H-)ead worm and that from the tail offspring as a (T-)ail worm.  We call the time between a worm’s birth (i.e. division of the parent worm) and its fission the reproductive waiting time (RWT) $\tau$, the worm size at division the Teilung size $s$, the worm size at birth the birth size, $A$, and the size accumulated during the RWT the added size, $\Delta$ (see Fig~\ref{Fig1}A). Because planarians are relatively two-dimensional, worm area is a reasonable approximation for worm size and significantly easier to obtain experimentally than worm volume. Worm sizes are measured in $mm^2$ and RWTs in units of days. The growth of the worm size between birth and division follows an exponential law with growth rate $\lambda$. We have imaged and recorded the above parameters for hundreds of worms of both species starting from a single worm establishing a clonal planarian family \cite{Carter2015}. Two clonal families for each species have been analyzed independently to account for within species variations.
\begin{figure}[!h]
\centering
\includegraphics[width=1.0\textwidth]{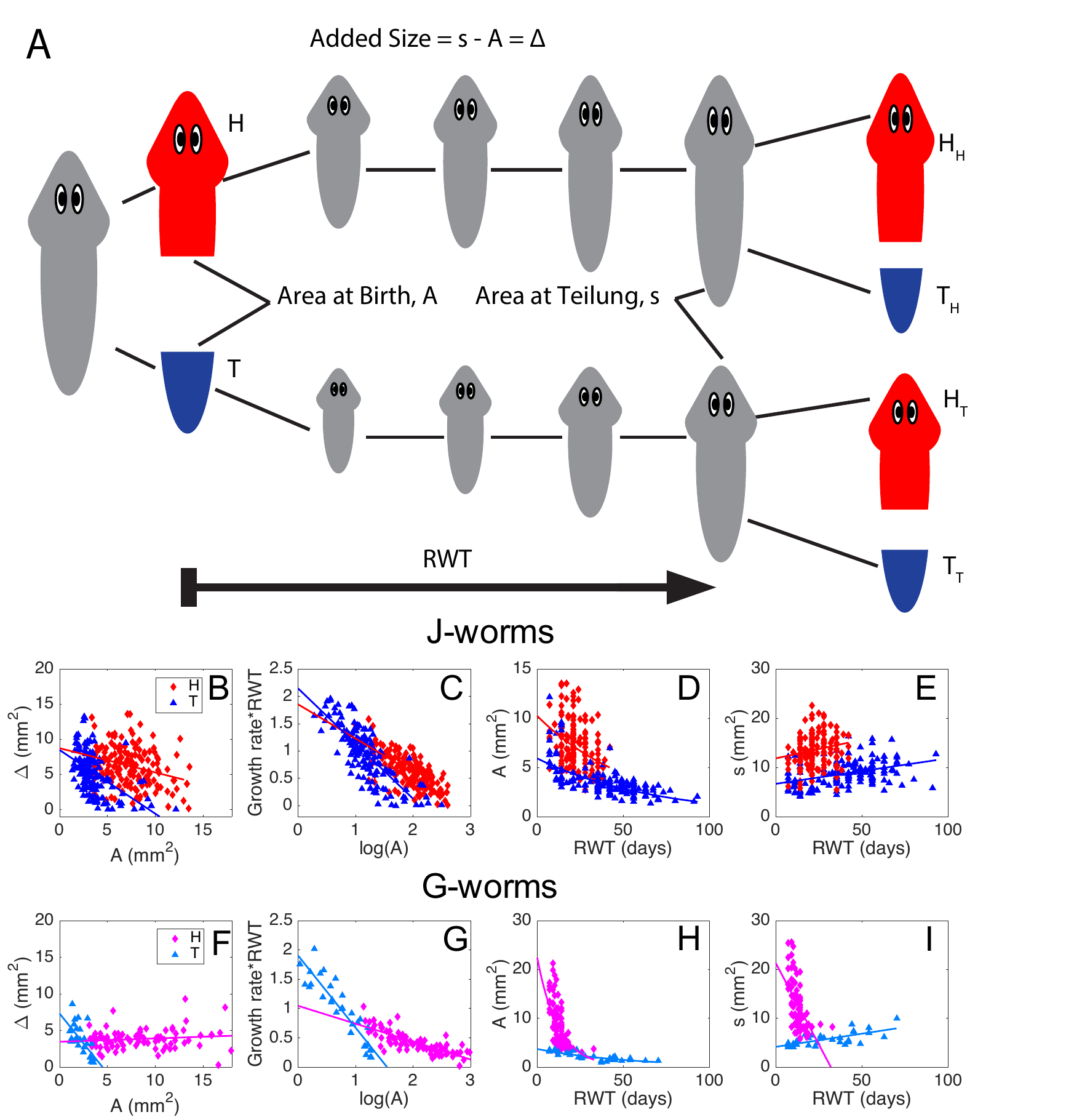}
\caption{{\bf Population dynamics: empirical relations between various reproductive variables imply complex reproductive behaviors of \textit{D. japonica} (J-worm) and \textit{D. tigrina} (G-worm) species.} (A) Schematic explaining key quantities. Experimental parameters for J-worms (B-E) and G-worms (F-I), respectively. H-worms are plotted in red and T-worms in blue. H and T denote H-worms and T-worms, respectively.}
\label{Fig1}
\end{figure}
\newpara
{\color{black}We start from the three basic size control strategies: sizer, adder, and timer. The perfect sizer and timer are characterized by a precise threshold of division size and RWT, respectively, at which the division occurs. The perfect adder is characterized by a constant added size, after which the division occurs. As reviewed in \cite{Araghi2015,Jun2015,Jun2016}, a plot of added size as a function of birth size is indicative of the reproductive strategy employed. A linear fit of the data would yield a slope of $-1$ for a perfect sizer where $\Delta = s-A$ with $s$ a constant, a slope of $0$ for a perfect adder, where $\Delta=C$ with $C$ a constant, and a slope of  $+1$  for a perfect timer where $\Delta = \lambda\tau A$ with $\lambda$ the growth rate and $\tau$ the constant RWT.

It's not the slope, however, which defines sizer, adder, or timer models. They are defined by the variables that cue the division, i.e. sizer, adder and timer are defined by the splitting rate function of Teilung size, added size and RWT, respectively. Perfect size control strategies correspond to a delta splitting rate function of the corresponding variable, while a general splitting rate function has a finite variance.
\newpara
Thus, general size control strategies could yield slopes different from the ideal cases, and are sometimes referred to as ``sloppy'' models \cite{Tyson1986}. A linear fit for the J-worm data yields a slope of $-0.34$ for H-worms and a slope of $-0.91$ for T-worms, neither of which fall into the perfect sizer, adder, or timer category (Fig~\ref{Fig1}B), but the T-worms favor a sizer mechanism. For G-worms, we obtain a slope of $0.046$ for H-worms, favoring adder mechanism. The slope of G T-worms is $-1.63$, which cannot be accounted for by a single model (Fig~\ref{Fig1}F). The observed deviations from the perfect slopes could also be due to mixed strategies such as the sizer-timer model in Ref. \cite{Osella2014}, where the splitting rate function depends on both Teilung size and RWT, or due to noise from finite data size. Thus, the slope alone is insufficient to determine the reproductive strategy. We therefore take the fitted slope in the $\Delta-A$ plot in Fig.\ref{Fig1} B and F as observables (among others) to infer the size control strategies in a more general inference scheme, as discussed below.

Similarly, the plot of $\lambda\tau$ as a function of $\log(A)$ is solely indicative of the reproductive strategy, where a slope of $-1$ indicates a perfect sizer with $\lambda\tau=\log(s)-\log(A)$ following from the exponential growth law, a slope of $0$ indicates a perfect timer with $\lambda\tau=C$, and a slope in between corresponds to a perfect adder. We thus take the slope of the $\lambda\tau-\log(A)$ plot (Fig.\ref{Fig1}C and G) as another observable for the inference. The slopes in the $\lambda\tau-\log(A)$ plot are consistent with those in the $\Delta-A$ plot for both J-worm and G-worm data.
\newpara
Finally, we plot birth size and Teilung size as a function of RWT in Fig.\ref{Fig1}(D,E,H,I) to extract the effective growth rate at a population level. The effective growth rate can be extracted as the slope from a linear fit of the $s-\tau$ plot (Fig.\ref{Fig1}E and I). $A-\tau$ plot is better fitted with an exponential, and we extract the slope between $\log(A)$ and $\tau$ as a supplementary observable to the effective growth rate. Additionally, for both species, H-worms show a significant spread in birth and Teilung areas, largely independent of their RWTs, whereas T-worms show a substantial spread in RWTs (Fig~\ref{Fig1}D-E, H-I). Interestingly, the G H-worms show the smallest area range, and upon fitting, yield a negative effective growth rate. This is due to the rapid division of G H-worms without time for substantial growth. We expect our model to capture this feature.}
\begin{figure}[!h]
\centering
\includegraphics[width=1.00\columnwidth]{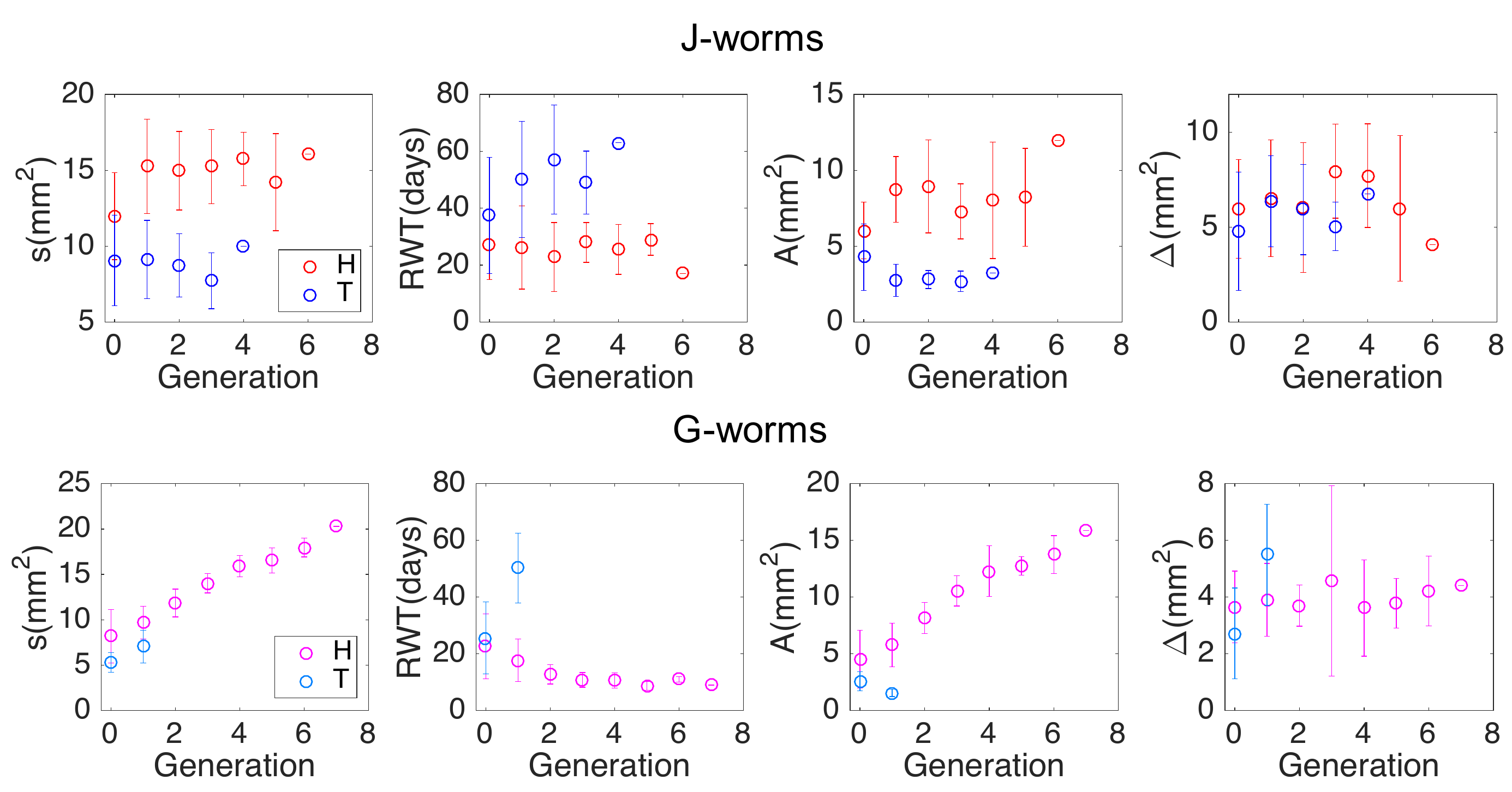}
\caption{{\bf Generation Dynamics: reproduction variables as a function of generation number for H-worm born from a H-parent (red) and T-worm born from a T-parent (blue).} (A-D): Teilung size, RWT, birth size and adde size as a function of generation number for J worms. (E-H) Teilung size, RWT, birth size and adde size as a function of generation number for G worms.}
\label{Fig2}
\end{figure}
\newpara
{\color{black}Fig.\ref{Fig1} summarizes the reproductive strategies at the population level. In parallel, we study the generation dynamics by plotting the mean and standard deviation of the birth size, Teilung size, added size and RWT as a function of generation number in Fig.\ref{Fig2}, by taking advantage of the lineage tracking of individual planarians. {\color{black} This means we follow the same-type offspring originating from a single individual, i.e. consecutive head and consecutive tail offspring lines.} We plot H-worms born from a H-parent (red) and T-worms born from a T-parent (blue), denoted as $H_H$ and $T_T$ respectively, with the exception that the zeroth generation worm is born from a parent of the opposite type ($H_T$ and $T_H$). This leads to a decoupling of the tail and head dynamics, and shows the distinctive behaviors of head and tail worms.
\newpara
The signatures of perfect sizer, adder, and timer models can be distinguished based on the generation dynamics. A perfect sizer would render all parameters generation-independent, provided that the worm follows a constant growth rate and divides with a constant proportionality. Generation dynamics of J-worms as shown in Fig.\ref{Fig2}A-D seem to favor such a sizer mechanism, but the variance is too large to be statistically significant.

A perfect adder model yields a generation-independent added size, but could render other parameters generation-dependent for a transient period when the worm sizes have not converged to the steady state value. G H-worms as shown in Fig.\ref{Fig2}E-H seem to favor such an adder mechanism, where the added size is insensitive to the generation number but the Teilung size and birth size increase with generation. This increase suggests that the added size is larger than the tail offspring the worm splits off. A steady state in size would be reached when the added size matches the tail offspring size provided that the worm divides with a constant proportionality.

A perfect timer will yield a constant RWT, but could render other parameters generation-dependent. It is worth noting that a timer model is equivalent to an adder model if the area growth is linear in time. With exponential growth, the worm size might diverge for a timer model, which is a biologically irrelevant regime. Similar to the population dynamics, the generation dynamics could be affected by mixed reproductive strategies, finite variance of the splitting rate function or finite size fluctuations. Therefore the generation data, together with the population data, needs to be incorporated into the general inference scheme that accounts for all these factors.
\newpara
We emphasize that the generation data is different from the population data, with the former comprised of worms following a specific lineage, and the latter involving all worms regardless of the lineage. Therefore the generation dynamics and the population dynamics will be treated as two independent observables to test the prediction of our model.

\begin{figure}[!h]
\includegraphics[width=1.00\columnwidth]{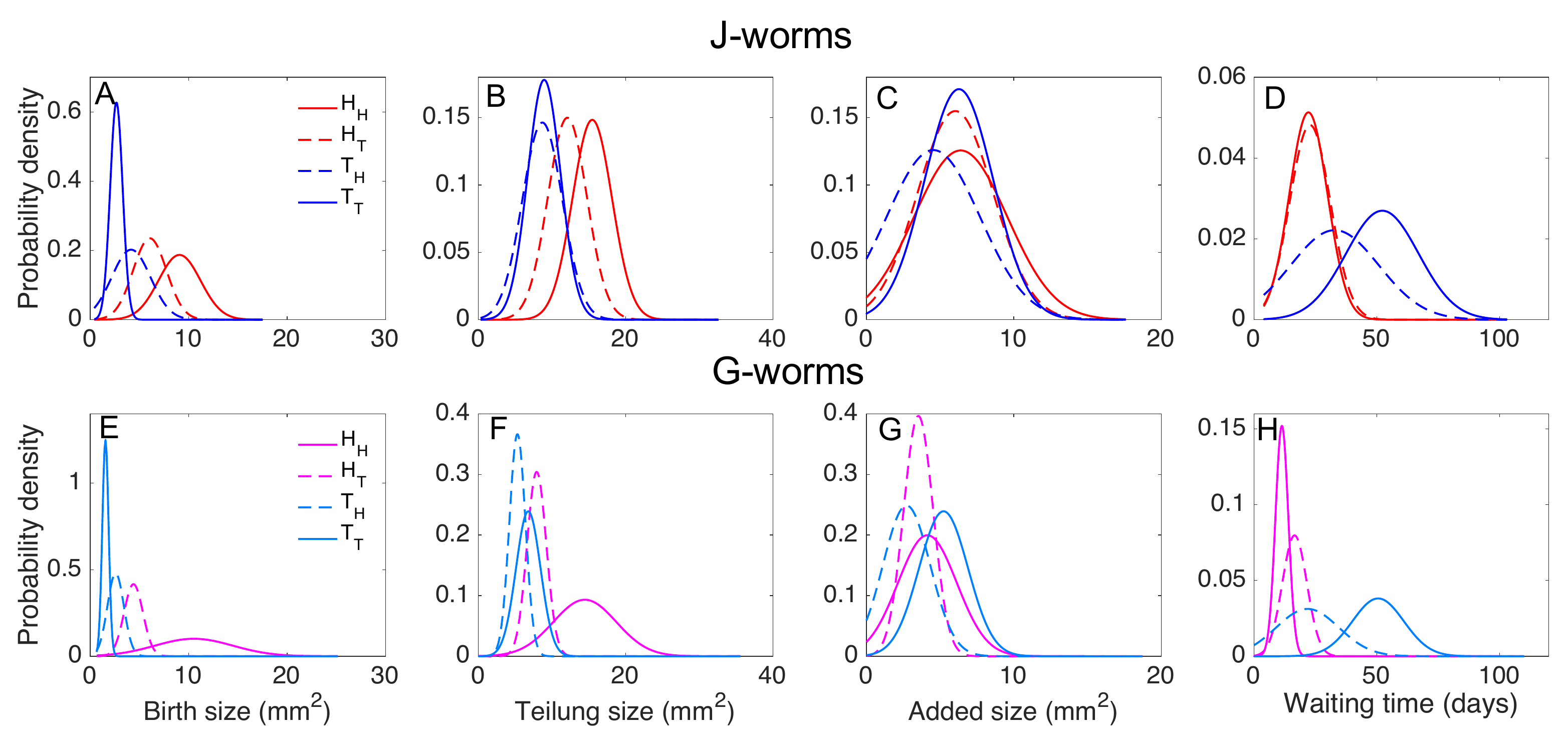}
\caption{{\bf Memory effect.} Gaussian-fitted (A) Birth size, (B) Teilung size, (C) added size, and (D) RWT distributions for H-worms born from H-parents ($H_H$), H-worms from T-parents ($H_T$), T-worms from H-parents ($T_H$), and T-worms from T-parents ($T_T$). Colors distinguish worm type, and line styles distinguish parent type. The splitted distributions encode a generational memory effect, mixed dynamics and asymmetry between T- and H-worms, rendering the four distributions to be generally distinct.}
\label{Fig3}
\end{figure}
\newpara
Additionally, the generation dynamics could be affected by a generational memory effect, which we have shown to be present in both species. The reproductive behavior of a planarian depends not only on its own identity, but also on its history, particularly the identity of its parent \cite{Dunkel2010,Carter2015}. To see this explicitly, we divide the worms into four groups, based on their own identity and the identity of their parent, categorized as H-worm from a head parent ($H_H$), T-worm from a tail parent ($T_T$), H-worm from a tail parent ($H_T$), and T-worm from a head parent ($T_H$). We neglect higher-order lineages due to the limited data size and the rapid decay of the memory effect with generation number \cite{Dunkel2010,Carter2015}. Distributions of Teilung size, birth size, added size and RWT of the four groups are plotted in Fig \ref{Fig3} for J- and G-worms. Worms of the same type but of different lineage (solid lines vs. dashed lines) show distinctive behaviors, suggesting the possibility of a memory effect. Of note, the memory effect is not the only factor to cause different distributions among the four groups. Intuitively, H-worms born from H-parents would have different birth areas than H-worms born from T-parents due to the distinctive Teilung sizes of H- and T-parents. Thus, even a pure memoryless sizer strategy would lead to different distributions of birth size and added size for worms of the same identity but different lineage. The memory effect enhances such differences among subgroups.}
\newpara
{\color{black}In summary, these data suggest that the two species employ different strategies to control their growth and division rates. Furthermore, within a species, heads and tails do not follow the same strategy either. The complicated behaviors of J H-worms and G T-worms cannot be captured by a single sizer, adder or timer mechanism, which calls for a mixed model that combines different mechanisms. The model has to account for the finite variance of the splitting rate function that deviates from the perfect size control mechanism and incorporate the generational memory present in both species. The parameters of the model need to be inferred using the population and the generation data to yield the optimal reproductive strategies for two worm types in both species. Moreover, the model has to be able to deal with limited data due to the slow reproductive cycles of the planarian.
\newpara
Below we present two models for the inference of the reproductive strategy, termed 2D model and additive model, both of which take the probability distributions of birth size, Teilung size, added size and RWT as inputs and use the population and generation data to infer the optimal reproductive strategy of each species. We show that the 2D model is limited by data size, but the additive model is not. The latter is able to straightforwardly quantify the proportions of sizerness, adderness or timerness within a mixed model and accurately captures the generation dynamics by incorporating the memory effect in both species.}

\subsection*{Mixed reproductive strategies: 2D model}
\begin{figure}[!h]
\begin{centering}
\includegraphics[width=1.00\columnwidth]{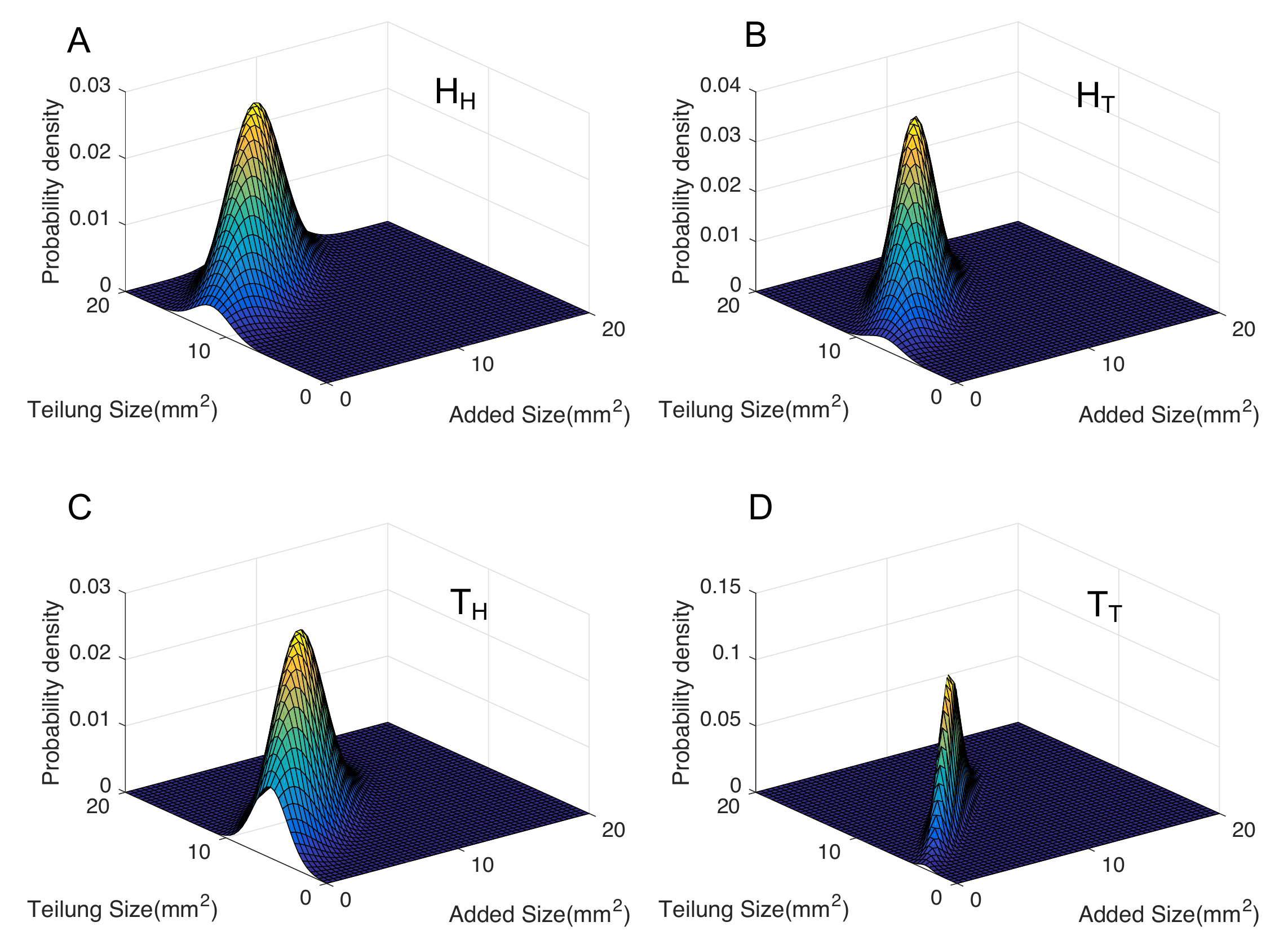}
\caption{{\bf 2D model} A-D: Gaussian-fitted joint probability distributions of Teilung size and added size for $H_H$, $H_T$ and $T_H$ and $T_T$ of J worms. The 2D model distinguishes from the additive model by incorporating the covariance, and is free of fitting parameters.}
\label{Fig4}
\end{centering}
\end{figure}
A straight forward generalization of the sizer, adder and timer model is a mixed 2D model with no free parameters, whose splitting rate function is obtained by integrating the 2D Teilung size-added size or Teilung size-RWT distributions (Fig.\ref{Fig4}), corresponding to the sizer-adder and sizer-timer model, respectively. It is therefore a direct representation of the multi-dimensional experimental data. We exclude the timer-adder model due to its apparent misfitting of the data as it predicts a positive slope in the added size \textit{vs.} birth size plot which is not observed in the data.

The memory effect is captured by assigning different splitting rate functions for worms from different groups, i.e. head worm from head parent ($H_H$), head worm from tail parent ($H_T$), tail worm from head parent ($T_H$) and tail worm from tail parent ($T_T$).
\newpara
The splitting rate functions for the sizer-adder and the sizer-timer 2D model are given by
\begin{eqnarray}
\label{eqn:gamma_2D_sizeradder}
\gamma(s,\Delta) = \frac{\rho(s,\Delta)}{1-\int_0^s ds'\int_0^{\Delta}d\Delta'\rho(s',\Delta')},\\
\label{eqn:gamma_2D_sizertimer}
\gamma(s,\tau) = \frac{\rho(s,\tau)}{1-\int_0^s ds'\int_0^{\Delta}d\tau'\rho(s',\tau')},
\end{eqnarray}
where $\rho(s,\Delta)$ and $\rho(s,\tau)$ are the 2D probability density distributions of the Teilung size-added size and Teilung size-RWT, respectively. To yield a continuous splitting rate function, as required by the simulation of the 2D model, a 2D Gaussian fitting is performed to the probability density distributions, which is plotted in Fig.\ref{Fig4}.
The division probability over the interval $[s,s+ds]$, $[\Delta,\Delta+d\Delta]$ for the sizer-adder model and $[s,s+ds]$, $[\tau,\tau+d\tau]$ for the sizer-timer model is obtained from the division probability
\begin{eqnarray}
\label{eqn:2D_division_sizeradder}
P(s\rightarrow s+ds,\Delta\rightarrow\Delta+d\Delta) = C\gamma(s,\Delta)dsd\Delta,\\
\label{eqn:2D_division_sizertimer}
P(s\rightarrow s+ds,\Delta\rightarrow\tau+d\tau) = C\gamma(s,\tau)dsd\tau.
\end{eqnarray}
The scaling factor $C$ is tuned to compensate for the low division probability at a high dimensional space along a path constrained by the relation between $s$ and $\Delta$ or $\tau$. In practice, it is chosen to match the output of the probability distributions from the simulation with the input distributions. In the case of the 2D sizer-adder model, we match the Teilung size and added size distributions, which yield $C=30$. Similarly, $C=80$ for the sizer-timer model by matching the Teilung size and the RWT distributions.
\newpara
Note that the 2D model has no free parameters, therefore it is a direct representation of the data. The population and generation dynamics data can be used to test the prediction of the sizer-adder and the sizer-timer model by calculating the corresponding likelihood. However, the 2D model requires the fitting of the variance and covariance of the input 2D Teilung size-added size and Teilung size-RWT distributions. The fitting of a 2D distribution generally requires more data than fitting a 1D distribution due to the finite covariance. The inference of the covariance requires more data than experimentally available, and therefore renders the 2D model data-limited. We perform a data sufficiency test below to illustrate this problem.

\begin{figure}[!h]
\begin{centering}
\includegraphics[width=1.00\columnwidth]{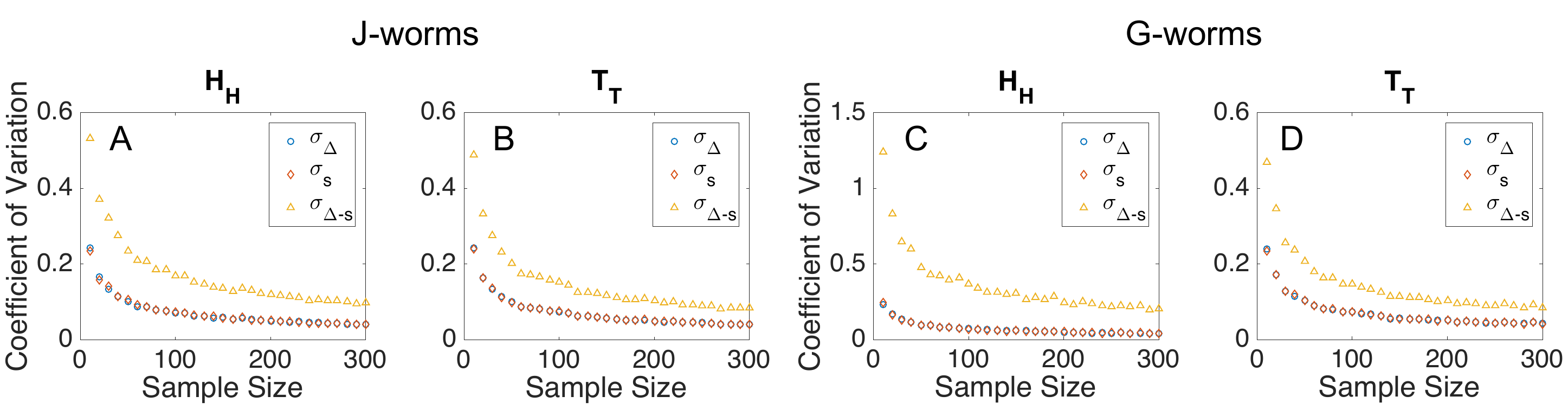}
\caption{{\bf Data sufficiency.} Data sufficiency check for J worms and G worms for the 2D model. The coefficient of variation (CV) of the distribution of inferred variance and covariance in Teilung size and added size are plotted as a function of sample size of the data. The covariance in the 2D model requires more data to be inferred accurately than what is available experimentally, rendering the 2D model data-limited.}
\label{Fig5}
\end{centering}
\end{figure}
Due to the slow reproduction cycle of planarians, the experimental data is size-limited with availability of several hundred worms per species. When divided into four groups to account for the memory effect, each group has at most $\sim100$ worms. Our model requires fitting the probability distributions of Teilung size, added size and RWT of the four groups to construct the splitting rate function. Assuming normal distribution, one needs to extract the mean, the variance and the covariance from the data.
\newpara
The data sufficiency problem can be formulated as: how many data points are required to infer the mean, the variance and the covariance of the normal distribution that generates the data to a specified accuracy? We perform numerical tests to answer this question. Data points are sampled with a 2D Teilung size-added size normal distribution based on the experimental data of J-worms and G-worms. A normal fit is performed on the sampled data to infer the variance and the covariance of the distributions that generate the data. We neglect the mean as it is much easier to fit than the variance. $600$ samplings are performed at each sample size to construct a distribution of the inferred variance and covariance, whose coefficient of variation (CV) is computed and plotted in Fig. \ref{Fig5}. The CV decreases as a function of sample size as expected. Choosing a threshold of $0.1$ for the CV, the 2D model requires more data than experimentally available to infer the covariance $\sigma_{\Delta-s}$ accurately. This renders the 2D model data-limited if we want to explicitly take generational memory into account, which, as we show below, is necessary to adequately capture the experimental data.
\newpara
To overcome the data-limitation and quantify each component within a mixed model, we developed a new additive model, which has the additional advantage over the 2D-model that it is able to explicitly specify the degree of sizerness, timerness, and adderness in each strategy for different species and worm types. The additive model outperforms the 2D model in capturing the population and generation dynamics and thus predicts the data that is not used to fit the splitting rate function better (Fig.\ref{Fig11} in SI).
\subsection*{Mixed reproductive strategies: Additive model}
%
The data sufficiency test in Fig.\ref{Fig5} shows that, in contrast to covariance, the variance of each dynamic variable can be fitted quite accurately with $\sim100$ data points, which is available experimentally. Therefore, instead of fitting the 2D probability density, we fit each 1D probability density independently as in Fig.\ref{Fig3}. Then we construct the 1D splitting rate function corresponding to the sizer, adder and timer component of the reproductive strategy
\begin{eqnarray}
\label{eqn:gamma_sizer}
\gamma_s(s)=\frac{\rho_s(s)}{1-\int_0^{s} \rho_s(s')ds'},\\
\label{eqn:gamma_adder}
\gamma_{\Delta}(\Delta)=\frac{\rho_{\Delta}(\Delta)}{1-\int_0^{\Delta} \rho_{\Delta}(\Delta')d\Delta'},\\
\label{eqn:gamma_timer}
\gamma_{\tau}(\tau)=\frac{\rho_{\tau}(\tau)}{1-\int_0^{\tau} \rho_{\tau}(\tau')d\tau'}.
\end{eqnarray}
\newpara
To construct a mixed model, we add the division probabilities of the sizer and adder or the sizer and timer component to form the sizer-adder and sizer-timer model. This additive approach is built on the simplifying assumption that both factors control reproduction independently and in parallel. Given our limited knowledge about the biological control of planarian reproduction, it is a reasonable possibility that two biological regulatory modules exist that cue the division independently and form an ``or gate'', enabling planarian division when either one of the modules is triggered. The total probability of division for the sizer-adder and sizer-timer model is thus given by
\begin{eqnarray}
\label{eqn:division_sizer_adder}
P(s\rightarrow s+ds,\Delta\rightarrow \Delta+d\Delta)=w \gamma_s(s)ds + (1-w) \gamma_{\Delta}(\Delta)d\Delta,\\
\label{eqn:division_sizer_timer}
P(s\rightarrow s+ds,\tau\rightarrow \tau+d\tau)=w \gamma_s(s)ds + (1-w) \gamma_{\tau}(\tau)d\tau.
\end{eqnarray}
A weight factor $w$ is introduced, which quantifies the weight of the sizer mechanism, or in short the ``sizerness''. The adderness and the timerness are quantified by $1-w$ provided that the total weights add up to one. This is another advantage of the additive model because the weight factor can be fitted to the population data (Fig.\ref{Fig1}) and the generation data (Fig.\ref{Fig2}) to yield the optimal mixed model with quantified weights of each component. It reveals how much each regulatory module controls the reproduction.
\newpara
We construct four division probabilities for the four groups ($H_H$, $H_T$, $T_H$ and  $T_T$) by integrating the corresponding probability densities (Fig.\ref{Fig3}) to incorporate the memory effect. Since the H-worm and the T-worm employ distinctive reproduction strategies, they are characterized by different weights $w$. We set $w=\alpha$ for the H-worm and $w=\beta$ for the T-worm, where $\alpha$ and $\beta$ are inferred by maximizing the likelihood of observing the population and the generation data. This is the standard maximum likelihood estimation technique (MLE).

We run simulations based on the division probability given in Eqs (\ref{eqn:division_sizer_adder})-(\ref{eqn:division_sizer_timer}) with $\alpha$ and $\beta$ ranging from $0$ to $1$ at an interval of $0.1$ for both the sizer-adder and the sizer-timer model. We define the log likelihood function as the sum of the log probability of observing the slopes from Fig \ref{Fig1} and the log probability of observing the mean Teilung size, added size, birth size, and RWT for the first generation worms in Fig.\ref{Fig2} given the model parameters $\alpha$ and $\beta$. We choose the first generation worms to provide equal number of data points between the generation data and the population data to avoid any bias in the total log likelihood:
\begin{equation}
\log(L) = \log[L_p] + \log[L_g],
\label{eqn:likelihood}
\end{equation}
where $L_p$ is the likelihood of observing the slopes and $L_g$ is the likelihood of observing the mean variables for the first generation worms. The detailed expressions for $L_p$ and $L_g$ are provided in the SI Text. The details of the simulation and the data analysis are provided in the Methods section.
\newpara
We plot the heat maps of the log likelihood, Eq (\ref{eqn:likelihood}), as a function of $\alpha$ and $\beta$ in Fig \ref{Fig6}(A-D). The optimal parameters of J H-worms indicates a mixture of 50\% sizer and 50\% adder with $\alpha=0.5$ in the sizer-adder model, while J T-worms are inferred to be a pure sizer with $\beta = 1.0$. The optimal parameter for G H-worms is $\alpha = 0.0$ with the sizer-adder model, indicating pure adder strategy. G T-worms have an optimal parameter $\beta = 0.4$, indicating 40\% sizer and 60\% adder. These results are consistent with the experimental data shown in Fig \ref{Fig1}-\ref{Fig2}. They pin down a specific mixed strategy with quantification of each contribution to the mixed dynamics for J H-worms and G T-worms, whose reproductive strategies are difficult to interpret from the empirical relations alone. The likelihood of the sizer-adder model for the G T-worms is relatively less sensitive to the sizerness $\beta$ (Fig.\ref{Fig6}C). This could be due to the smaller data set for G T-worms, which was experimentally limited to $\sim 30$ worms per family, because G-worm tails experienced a comparably high death rate~\cite{Carter2015}.
\newpara
It is worth noting that the optimal sizer-adder strategy and the optimal sizer-timer strategy yields similar performance based on the likelihood, which is consistent with the observation for \textit{E.coli}~\cite{Osella2014} and a test with artificially generated test data (see Fig.\ref{Fig7} in SI). The optimal sizer-timer model for J H-worms corresponds to $\alpha=0.8$, indicating 20\% timer and 80\% sizer. This suggests that a mixture of sizer-timer strategy can yield similar population and generation dynamics as the sizer-adder strategy by adjusting the relative proportion of sizerness. This is reasonable as we know that the timer model is exactly equivalent to the adder model under a linear growth law. With an exponential growth for the planarians, the exact mapping no longer holds, but given that the growth rate is small, the timer mechanism is still similar to the adder mechanism for worms with moderate size. Because a pure timer mechanism can lead to size divergence, the timer component in the mixed strategy has to be constrained within a threshold as illustrated in Fig.\ref{Fig6}B and D. Similarly, for G H-worms, which are best described by an adder mechanism, the corresponding optimal sizer-timer strategy yields $\alpha=0.7$, indicating a mixture of 30\% timer and 70\% sizer. This suggests that a mixed sizer-timer strategy can have a similar performance as a pure adder strategy. This is in agreement with two independent works proposing adder \cite{Araghi2015} and sizer-timer strategies \cite{Osella2014} for \textit{E.coli}, respectively.
\begin{figure}[!h]
\includegraphics[width=1.00\columnwidth]{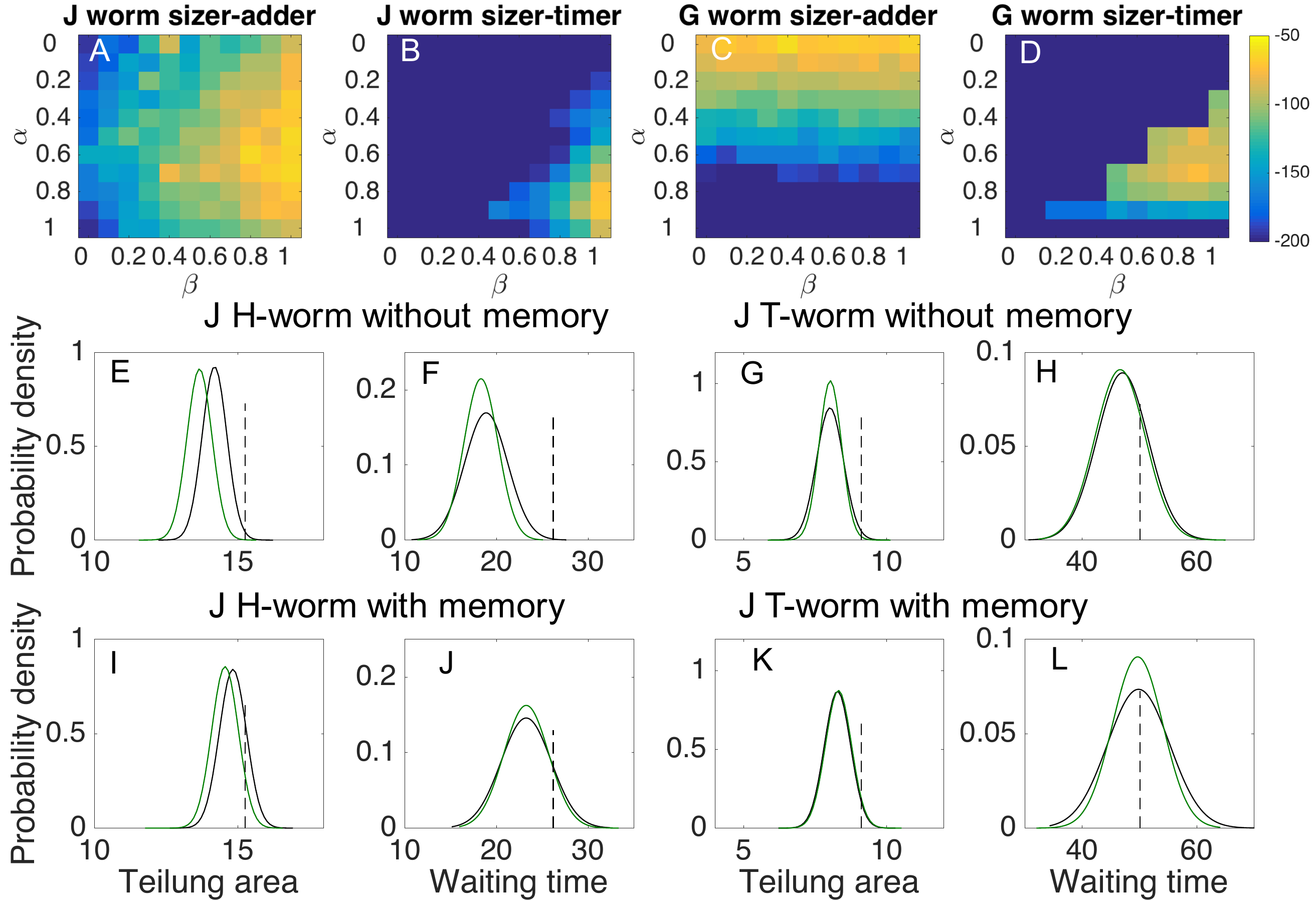}
\caption{{\bf Inferred reproductive strategies and generational memory.} A-D: Heat maps for the log likelihood of the additive model with memory in the parameter space of $\alpha$ (sizerness for head worms) and $\beta$ (sizerness for tail worms). The model identifies the J T-worm to be a pure sizer, while the G H-worm is a pure adder. The J H-worm and the G T-worm implement mixed reproductive strategies and can be captured by sizer-adder and sizer-timer models equally well. E-H (I-L): Distributions of mean Teilung size and mean RWT of the first generation worms ($H_{HT}$ and $T_{TH}$) from simulations using the optimal sizer-adder model (black) and sizer-timer model (green) without (with) memory. The dashed vertical lines denote the experimental data. The incorporation of memory effect improves the prediction of the generation data.}
\label{Fig6}
\end{figure}
\newpara
To check the significance of the memory effect explicitly incorporated into the additive model, we construct a ``memoryless model'' where the splitting rate function does not depend on the identity of the parent worm. This is realized by dividing the data into only two groups for H-worms and T-worms. We integrate the probability density of the Teilung size, added size and RWT for these two groups to construct the corresponding splitting rate function and repeat the analysis by plotting the heat maps of the log likelihood (Fig.\ref{Fig8} in SI). We find that the overall likelihood of both sizer-adder and sizer-timer model has significantly dropped. The drop is more pronounced for the sizer-timer model. A careful study reveals that this is due to the failure of the memoryless model to capture the generation data (Fig.\ref{Fig6} E-H for J-worms and Fig.\ref{Fig8} E-H in the SI for G-worms). The explicit incorporation of the memory effect thus significantly improves the prediction of the generation dynamics (Fig.\ref{Fig6} I-L for J-worms and Fig.\ref{Fig8} I-L in the SI for G-worms). We emphasize that the improvement of the prediction for the statistics of the first generation worms is not a direct consequence of the input distributions from Fig.\ref{Fig3}, but a true prediction of the RWT for the sizer-adder model, which is entirely due to the memory effect as RWT is not an input. The improvements can be seen in both H- and T-worms, but are more significant in the H-worms (T-worms) for J-worms (G-worms). Furthermore, the improvements are robust against numerical fluctuations (Fig.\ref{Fig10} in SI).
\newpara
Taken together, we have shown that our simplified approach based on an additive model - in contrast to the full 2D model - successfully captures the reproductive strategies of asexually reproducing multi-cellular organisms (planarians) under the constraint of limited data availability due to long reproduction cycles and generational memory. Furthermore, the additive model quantifies the contribution of each size control strategy and provides new insights into how generational memory affects asexual reproduction of the two species of freshwater planarians studied here.

\section*{Biological interpretation}
The data presented here and in Ref.~\cite{Carter2015} suggest that J-worms optimize the survival and reproductive success of both offspring at the cost of comparably longer RWTs, whereas G-worms optimize the frequency of division and the reproductive success of the H-worm at the cost of increased tail death. On the individual worm level, this manifests through different size control strategies for the H- and T-offspring for both planarian species.
\newpara
In the case of J-planarians, T-worms are best described as pure sizers (Fig.\ref{Fig6}A), where the division is controlled entirely by an individual's size. Once a critical size is reached, division occurs. A pure sizer mechanism ensures that fission does not occur while the planarian is too small to produce two viable offspring. In contrast, J H-worms fit best within a sizer-adder or sizer-timer model (Fig.\ref{Fig6}A-B). Thus, while size still matters, a second mechanism exists which co-controls the reproductive dynamics.
\newpara
In the case of G-planarians, H-worms are always born sufficiently big to divide (Fig.\ref{Fig1}F) and thus are not regulated by absolute size. They act as pure adders by regenerating a new tail of minimal size and dropping it off immediately, thus optimizing the reproduction rate and the resource allocation to the head piece at the cost of making small tails. Why do G T-worms exhibit a mixed strategy and are not pure sizers like J T-worms? We caution that the data is not sufficient to rule out the possibility of a pure sizer mechanism for G T-worms, which have an extremely negative slope in Fig.\ref{Fig1}F, larger than that expected for even a pure sizer. This extremely negative slope could be due to the small sample size or indicative of a strategy outside the current framework.
\newpara
Since little is known about the molecular mechanisms regulating planarian asexual reproduction and growth, the biological interpretation of the observed strategies is far from trivial. Anatomically, there are three main differences between heads and tails that are shared in both species. H-worms are generally born bigger (Fig.\ref{Fig1}B,F) and inherit the brain and eating machinery (pharynx). Given that the respective H- and T-worms of the two species do not share the same strategy suggests that neither birth size nor the head/tail distinction are sufficient to explain the different strategies. This may be due to the fact that both pharynx and brain regeneration are completed within 5-6 days~\cite{Kobayashi1999} and 7-12 days~\cite{Hagstrom2015, Cebria2007, Agata2008}, respectively, which is sufficiently short when compared to T-worm RWTs. Thus, T-worms possess a pharynx and a brain for most of their RWT. Of note, the brain has been previously implicated in fission control in the planarian species \textit{Dugesia dorotocephala}, where it was shown that fission frequency was influenced by environmental photoperiods, mediated through melatonin release in the brain~\cite{Morita1984}. Because planarians exposed to exogenous melatonin in their aquatic environment exhibited decreased fission rates~\cite{Morita1984}, whereas decapitated planarians fissioned more frequently \cite{Brondsted1955, Hori1998, Morita1984}, high melatonin levels appear to suppress fission. How this suppression works is unknown. It is also unclear how it fits into our framework of size control mechanisms, since we do not know whether melatonin affects H-worms and T-worms the same and how melatonin affects fission frequency as a function of planarian size. Systematic studies of these relationships would constitute a reasonable first step toward a mechanistic understanding of size control.
\newpara
In summary, we find that the employed size control mechanisms of individual planarians revealed in this study can explain some of the features of the reproductive strategies \cite{Carter2015} of the two planarian species investigated. Our findings can serve as a point of departure for future studies on the molecular mechanisms governing planarian asexual reproduction.

\section*{Conclusion}

We have quantitatively studied the size-control and asexual reproduction strategies of two freshwater planarian species, \textit{D. japonica} and \textit{D. tigrina}. Taking advantage of a unique experimental dataset, we constructed a novel additive model that mixes various reproductive strategies based on worm sizes, added sizes and reproductive waiting times. In contrast to the 2D model, which is a parameter-free representation of the data and suffers from data insufficiency, the additive model can be fit to small datasets and explicitly quantifies the proportions of each strategy in the mixed dynamics. Based on the likelihood of reproducing the experimental data through simulation with the additive model, we have identified distinct reproduction strategies for both species and showed that they are influenced by generational memory. The incorporation of the memory effect into the additive model successfully captures the generation dynamics. Aside from our ability to infer size control strategies of multi-cellular organisms for the first time, the diversity of approaches we find utilized by species within the same genus and even offspring (H/T) of the same species, is surprising. It will be interesting to dissect which molecules and pathways underlie these different control strategies and what evolutionary advantage they pose. More generally, our theoretical approach will be useful for inferring size control strategies of other single or multi-cellular fissiparous organisms where data is limited as well as for gaining insights into the relative contributions of the three strategies to help decipher the underlying molecular control mechanisms.

\section*{Acknowledgments}
This project was supported by the BWF CASI and NSF CAREER 1555109 award (to EMSC), NIH K25 GM098875-02 (to DJS), and the URS Hellman Physics award (to KJK). This research was supported in part through computational resources provided by Syracuse University. The authors thank A. Tran, Y. He, and M.P. Truong for help with planarian care and imaging, M. Vergassola, O. Cochet-Escartin, and R. Wang for comments on the manuscript.

\nolinenumbers

\section*{Supporting Information}


%
\subsection*{1. Comparison between the 2D model and the additive model}

We show in this section that the additive model outperforms the 2D model with generational memory. We integrate the Gaussian-fitted Teilung size-added size and Teilung size-RWT distributions (Fig.4 in the main text) to obtain the splitting rate functions (Eqn.2-3 in the main text) for the 2D sizer-adder and sizer-timer model, respectively. We perform the simulation with the splitting rate functions and record the population and generation dynamics. $500$ simulations are run for each model and a log likelihood is calculated based on the generation and population dynamics of the H- and T-worms (Eqn. \ref{eqn:likelihood_SI}-\ref{eqn:likelihood_g_SI}).
\newpara
A similar procedure is followed for the additive model, but with 1D splitting rate functions constructed from Teilung size, added size and RWT distributions independently. We add up the division probabilities from sizer and timer, or sizer and adder, to yield the additive sizer-timer and sizer-adder model, respectively. A weight factor determining the relative sizerness in the mixed model is fitted using the population and generation data. We identify the optimal additive sizer-adder model with $\alpha=0.5$, $\beta=1.0$ and sizer-timer model with $\alpha=0.8$, $\beta=1.0$, corresponding to a pure sizer mechanism for T-worms and a mixed strategy for H-worms. We plot the log likelihood of the 2D model and the optimal additive model fitted with the population and the generation data in Fig.\ref{Fig11}. Both models have incorporated generational memory. The additive model outperforms the 2D model for both the sizer-adder and the sizer-timer strategy. This suggests that the additive model works better than the 2D model with limited data and quantifies each component within a mixed reproductive strategy.
\newpara
The reason why the additive model outperforms the 2D model is two-fold. Firstly, the 2D model takes the 2D distributions of Teilung size-added size or the Teilung-size RWT as the input, which requires a multivariable Gaussian fitting of the distributions. As demonstrated in Fig. 5 in the main text, the fitting of a 2D Gaussian distribution requires more data than what is experimentally available. Therefore, an overfitting occurs when we fit the 2D Gaussian distribution to a limited data set. This renders the 2D model less predictive when it comes to the population and generation dynamics. The additive model does not suffer from this issue as it only requires the fitting of two 1D Gaussian distributions on each variable independently. Secondly, the additive model has a free parameter that quantifies the sizerness in the mixed strategy, which can be tuned to capture the generation and population dynamics to yield the optimal additive model. In contrast, the 2D model does not have any free parameters. This extra degree of freedom allows the additive model to use more information in the data to optimize its performance. Essentially, the additive model uses the population and generation dynamics to determine the relative weight of each size control mechanism in the mixed strategy, which is unattainable in the 2D model.
\begin{figure}[!h]
\begin{centering}
\includegraphics[width=1.00\columnwidth]{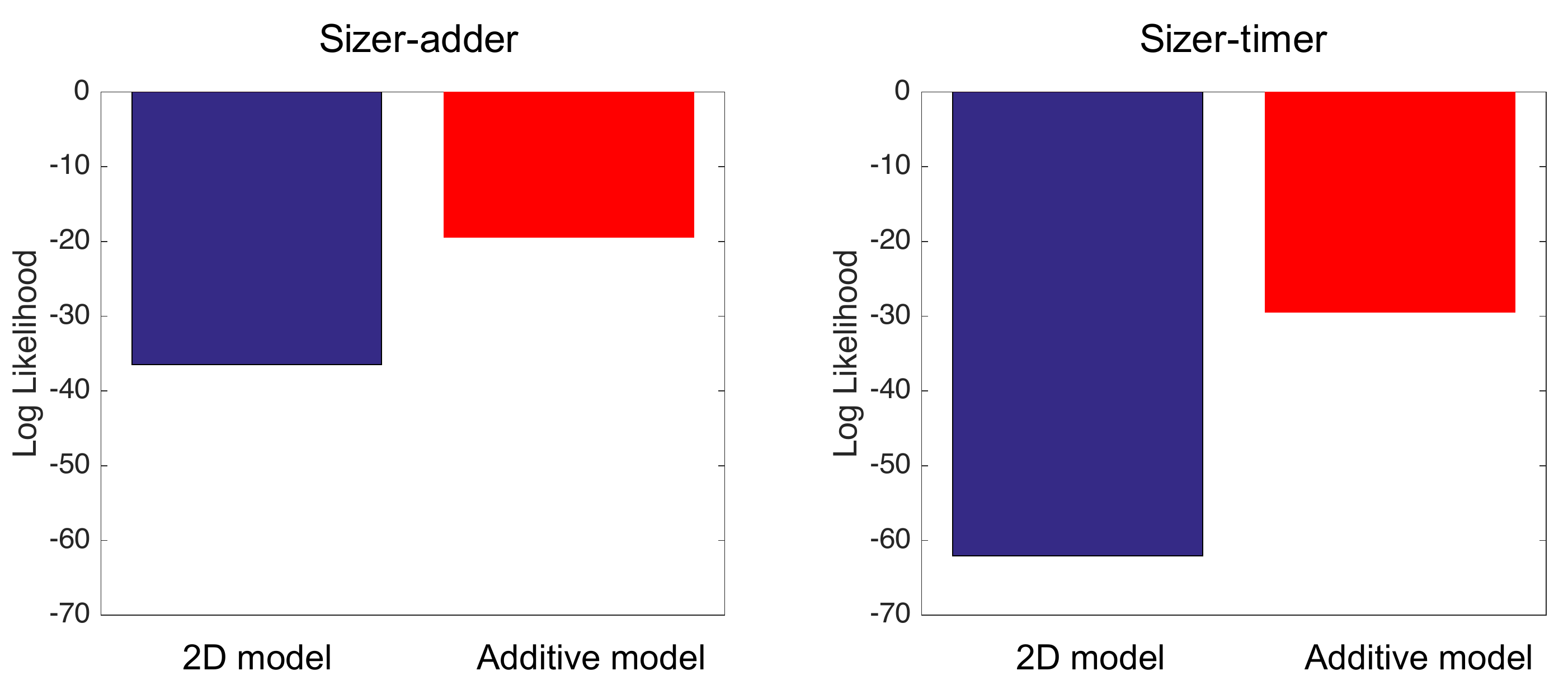}
\caption{{\bf 2D model vs. additive model for J worm.} Blue (red) bars correspond to the log likelihood of the 2D model (additive model) with generational memory. The additive model outperforms the 2D model.}
\label{Fig11}
\end{centering}
\end{figure}
\subsection*{2. Consistency of additive model and the parameter inference}
The additive model uses the population and generation data to infer the sizerness in the mixed strategy as parameterized by $\alpha$ and $\beta$ for the H- and T-worms, respectively. We check the consistency of the additive model and the parameter inference scheme using artificial data generated from the simulation of the additive sizer-adder model with $\alpha = 0.5$ $\beta = 1.0$ and the additive sizer-timer model with $\alpha = 0.8$ $\beta = 1.0$, corresponding to the optimal parameters from the experimental data for J-worms. The goal is to verify that $\alpha$ and $\beta$ can be inferred correctly by fitting the population and generation data using maximum likelihood estimation (MLE). The artificial data has the same format and size as the experimental data with Teilung size, birth size, added size, RWT, growth rate, division ratio and lineage recorded for each planarian. We extract splitting rate functions from the artificial data and run 100 simulations for each pair of $\alpha$ and $\beta$ to plot the log likelihood of the parameters given the population and generation dynamics as a function of $\alpha$ and $\beta$. To check if the sizer-adder model can be distinguished from the sizer-timer model, we perform the inference using each model on data generated by both models. The results are summarized in Fig.~\ref{Fig7}. We recover the correct $\alpha$ and $\beta$ by maximizing the log likelihood, validating the consistency of the model and the inference approach. However, based on the log likelihood, the optimal sizer-adder model cannot be distinguished from the optimal sizer-timer model as the optimal parameters in both models fit the artificial data equally well, which is consistent with the observation from \textit{E.coli} reproduction \cite{Osella2014}.
\begin{figure}[!h]
\begin{centering}
\includegraphics[width=1.00\columnwidth]{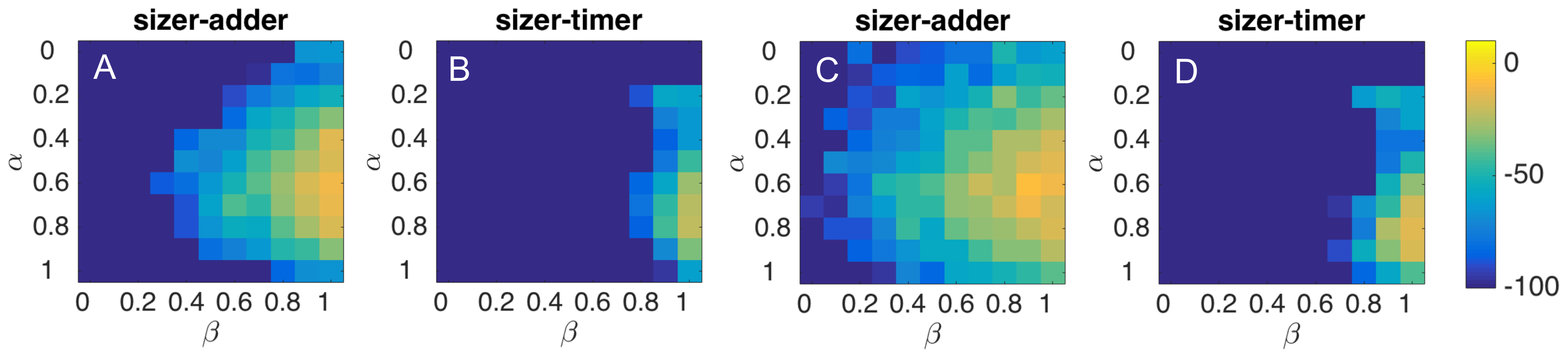}
\caption{{\bf Inference with artificial data generated from simulations.} Log likelihood of the additive model based on artificial data generated from the simulation. Two sets of data are generated using the sizer-adder model with $\alpha=0.5$ $\beta=1.0$ (A-B) and the sizer-timer model with $\alpha=0.8$ $\beta=1.0$ (C-D), respectively. Both data sets are used as input of the sizer-adder and sizer-timer model to produce the heat map of the log likelihood based on the population and the generation data in the parameter space of $\alpha$ and $\beta$. The additive model predicts the parameters that generate the data correctly through maximum likelihood estimation, validating the consistency of the model and the inference scheme. However, it also renders sizer-adder and sizer-timer models indistinguishable as the optimal parameters in both models fit the artificial data equally well.}
\label{Fig7}
\end{centering}
\end{figure}
\subsection*{3. Maximum likelihood estimation (MLE)}
The optimal model parameters $\alpha$ and $\beta$ that quantify the sizerness of H- and T-worms in the additive model are determined by maximizing the log likelihood given the experimental population and the generation data. The log likelihood of $\alpha$ and $\beta$ can therefore be decomposed into the population part $log(L_p)$ and the generation part $log(L_g)$ as
\begin{eqnarray}
\label{eqn:likelihood_SI}
\log(L) = \log[L_p] + \log[L_g],\\
\label{eqn:likelihood_p_SI}
\log[L_p(\alpha,\beta|a,b,c,d)] = \sum_{i\in{H,T}} \{\log[p^i_{\Delta,A}(a|\alpha,\beta)] + \log[p^i_{\log(A),\lambda\tau}(b|\alpha,\beta)]\\\nonumber + \log[p^i_{\log(A),\tau}(c|\alpha,\beta)]  +  \log[p^i_{s,\tau}(d|\alpha,\beta)]\},\\
\label{eqn:likelihood_g_SI}
\log[L_g(\alpha,\beta|\Delta,A,s,\tau)] = \sum_{i\in{H,T}} \{\log[p^i_{\Delta}(\Delta|\alpha,\beta)]+ \log[p^i_{A}(A|\alpha,\beta)]\\\nonumber +  \log[p^i_{s}(s|\alpha,\beta)] + \log[p^i_{\tau}(\tau|\alpha,\beta)]\},
\end{eqnarray}
where $p$ are the probability densities of the fitted slopes, denoted by a-d (Fig.1 in the main text), and the mean reproductive variables for the first generation worms ($\Delta$, $A$, $s$ and $\tau$) (Fig.2 in the main text) at given $\alpha$ and $\beta$, which correspond to the population and generation data, respectively. The probability densities are obtained from 500 simulations for each pair of $\alpha$ and $\beta$. We assume that $a-d$, $\Delta$, $A$, $s$ and $\tau$ are all independent observables, which justifies the summing of the log probabilities of each individual observable to yield the total log likelihood. This assumption is justified by the consistency check with artificial data as demonstrated in section 2.
\subsection*{4. Memoryless model vs memory model}
In the main text, we studied the additive model with generational memory, here we study the ``memoryless model'' for comparison and show that the memory effect has to be incorporated to capture the generation dynamics. In the memoryless model, the reproduction dynamics depend only on a planarian's identity, i.e. whether it is a H-worm or a T-worm, regardless of the identity of its parent. This is realized by dividing the data into only two groups for H-worms and T-worms, and integrating the probability distributions of the Teilung size, added size and RWT for these two groups to construct the corresponding splitting rate function. We repeat the same analysis as we did for the memory model by plotting the heat maps of the log likelihood of $\alpha$ and $\beta$ given the population and the generation data in Fig.\ref{Fig8}. The overall likelihood of the memoryless model is much smaller than the memory model, suggesting that the absence of the memory effect leads to a poor prediction of the population and generation dynamics. Furthermore, the optimal sizer-adder strategy significantly outperformed the optimal sizer-timer strategy in this framework. Since the two optimal mixed strategies are expected to perform equally well, as previously shown for \textit{E.coli} \cite{Osella2014} and demonstrated using artificial data (Fig.\ref{Fig7}), these results suggest that the data cannot be entirely captured by the memoryless model.
\newpara
To pin down which part of the data the memoryless model fails to capture, we decompose the total log likelihood into the population part and the generation part (Eqn.\ref{eqn:likelihood_SI}-\ref{eqn:likelihood_g_SI}). The population dynamics is equally well-captured by the optimal sizer-adder and the optimal sizer-timer models. However, both models fail to capture the generation data. To see this explicitly, we plot the distributions of the mean birth size, Teilung size, added size, and RWT for the first generation J-worms (Fig. 6 in the main text) and G-worms (Fig.\ref{Fig8}). The distributions are obtained from simulations with the optimal memoryless sizer-adder and sizer-timer model. Both memoryless models fail to capture the experimental data as denoted by the dashed vertical line, but the memoryless sizer-adder model performs slightly better than the memoryless sizer-timer model at the tail of the distributions for J H-worms (Fig.6 E-F) and G T-worms (Fig.\ref{Fig8} G-H). The latter explains the bias toward the sizer-adder model over the sizer-timer model. The poor prediction of the generation dynamics by the memoryless model suggests that the generational memory effect, where the reproduction dynamics depend not only on the identity of the reproducing worm but also on its lineage, needs to be taken into account. The incorporation of the memory effect significantly improves the prediction of the generation dynamics and renders the mixed sizer-adder model indistinguishable from the sizer-timer model as expected.
\begin{figure}[!h]
\includegraphics[width=1.00\columnwidth]{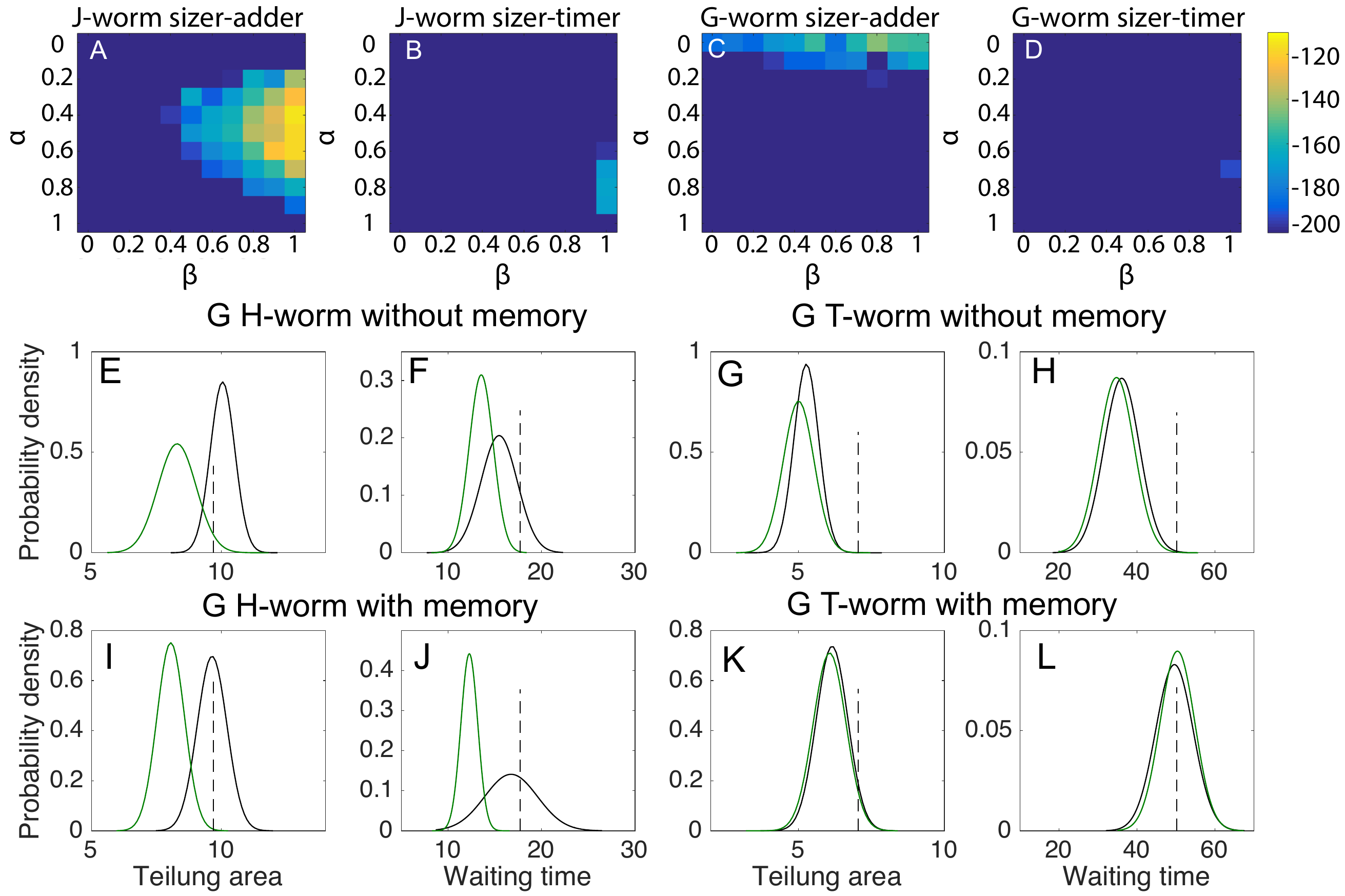}
\caption{{\bf Memoryless vs. memory model} A-D: Heat maps for the log likelihood of the additive model without memory (A) J-worm sizer-adder, (B) J-worm sizer-timer, (C) G-worm sizer-adder, and (D) G-worm sizer-timer. $\alpha$ is the sizerness for H-worms and $\beta$ is the sizerness for T-worms. E-H (I-L): Distributions of mean Teilung size and mean RWT of the first generation worms ($H_{HT}$ and $T_{TH}$) from simulations with the optimal sizer-adder model (black) and sizer-timer model (green) without (with) memory. The dashed vertical lines denote the experimental data. The incorporation of generational memory improves the prediction of the generation data.}
\label{Fig8}
\end{figure}
{\color{black}\subsection*{5. Numerical fluctuations in the simulation due to finite data size}}
\begin{figure}[!h]
\begin{centering}
\includegraphics[width=1.00\columnwidth]{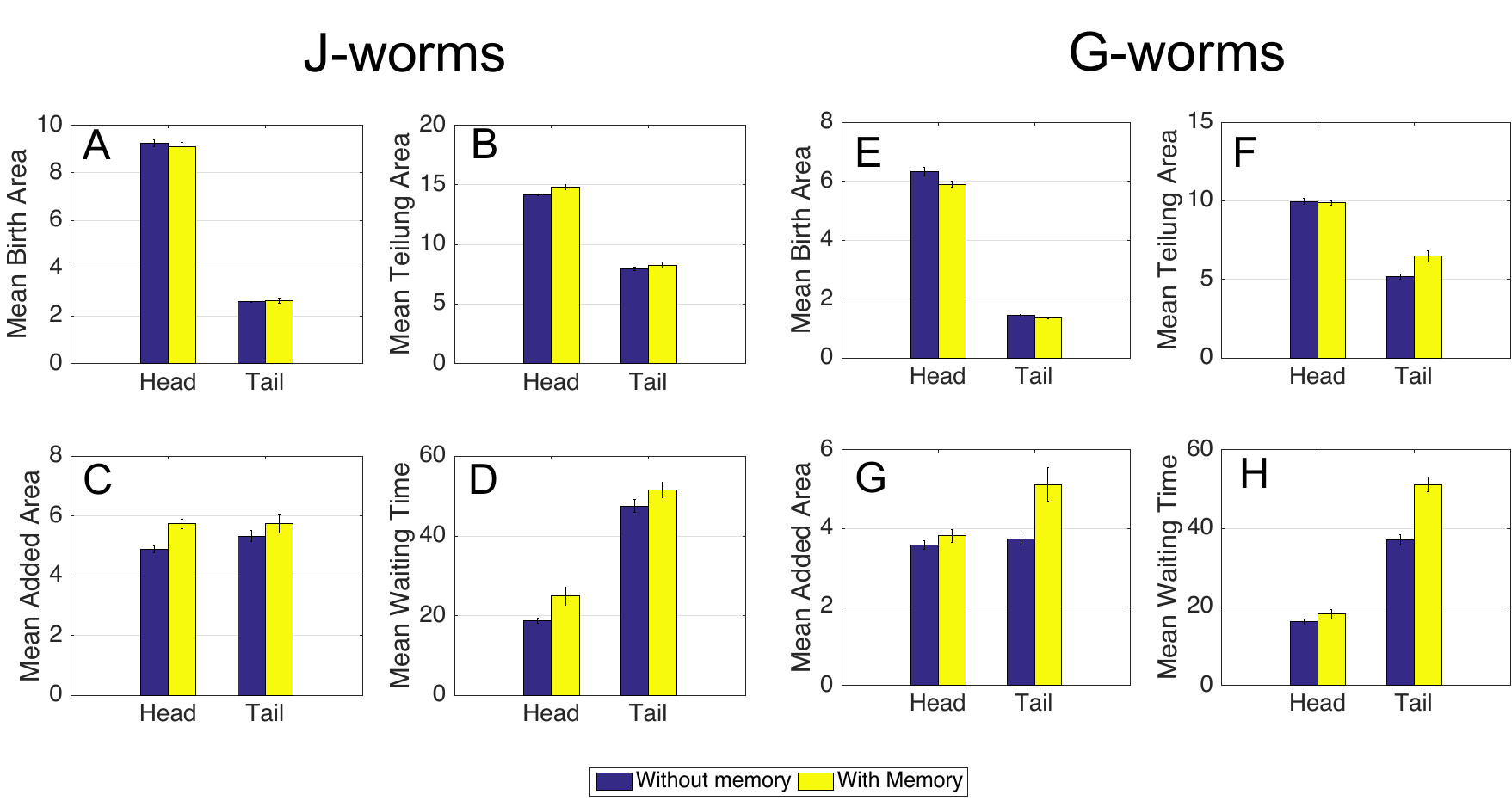}
\caption{{\bf Numerical fluctuations in the simulation.} Numerical fluctuations in the simulation that produces the distributions of the mean Teilung size, added size, birth size and RWT in the generation dynamics for J-worms and G-worms. The bar plots are the mean of the 5 distributions of mean Teilung size, added size, birth size and RWT. The error bars are the standard deviation of the mean, representing the numerical fluctuations of the simulation due to finite size effect.}
\label{Fig10}
\end{centering}
\end{figure}
We have shown that the memory effect improves the prediction of the generation dynamics by shifting the distributions of mean sizes and RWTs for the first generation worms. The distributions are produced from 100 simulations and are subject to numerical fluctuations due to the finite data size. To estimate the magnitude of the fluctuations, we perform 500 simulations with and without memory effect each and divide the data into 5 groups to produce 5 sets of distributions of mean Teilung size, added size, birth size and RWT. We plot the mean of the distributions as a bar plot overlaid with error bars representing the numerical fluctuations of the mean across the 5 distributions in Fig.\ref{Fig10}. We conclude that the numerical fluctuations are negligible as compared to the shift of the distribution due to the memory effect. Therefore, the improvement of the prediction of the generation dynamics due to the memory effect is robust.

\end{document}